\documentclass[10pt,twocolumn,letterpaper]{article}

\usepackage{cvpr}
\usepackage{times}
\usepackage{epsfig}
\usepackage{graphicx}
\usepackage{amsmath}
\usepackage{amssymb}

\usepackage[breaklinks=true,bookmarks=false]{hyperref}

\cvprfinalcopy 

\def\cvprPaperID{****} 

\begin{document}


\title{RCLC: ROI-based joint conventional and learning video compression}

\author{Trinh Man Hoang\\
Graduate School of Science and Engineering, Hosei University\\
Tokyo, Japan\\
{\tt\small trinhman.hoang.3f@stu.hosei.ac.jp}

\and
Jinjia Zhou\\
Graduate School of Science and Engineering, Hosei University\\
JST, PRESTO, Tokyo, Japan\\
{\tt\small zhou@hosei.ac.jp}
}
\maketitle

\begin{abstract}
   COVID-19 leads to the high demand for remote interactive systems ever seen. One of the key elements of these systems is video streaming, which requires a very high network bandwidth due to its specific real-time demand, especially with high-resolution video. Existing video compression methods are struggling in the trade-off between the video quality and the speed requirement. Addressed that the background information rarely changes in most remote meeting cases, we introduce a Region-Of-Interests (ROI) based video compression framework (named RCLC) that leverages the cutting-edge learning-based and conventional technologies. In  RCLC, each coming frame is marked as background-updating (BU) or ROI-updating (RU) frame. By applying the conventional video codec, the BU frame is compressed with low-quality and high-compression, while the ROI from RU-frame is compressed with high-quality and low-compression. The learning-based methods are applied to detect the ROI, blend background-ROI, and enhance video quality. The experimental results show that our RCLC can reduce up to 32.55\% BD-rate for the ROI region compared to H.265 video codec under a similar compression time with 1080p resolution.
\end{abstract}

\section{Introduction}
\begin{figure*}[t]
\begin{center}
\includegraphics[width=0.8\linewidth]{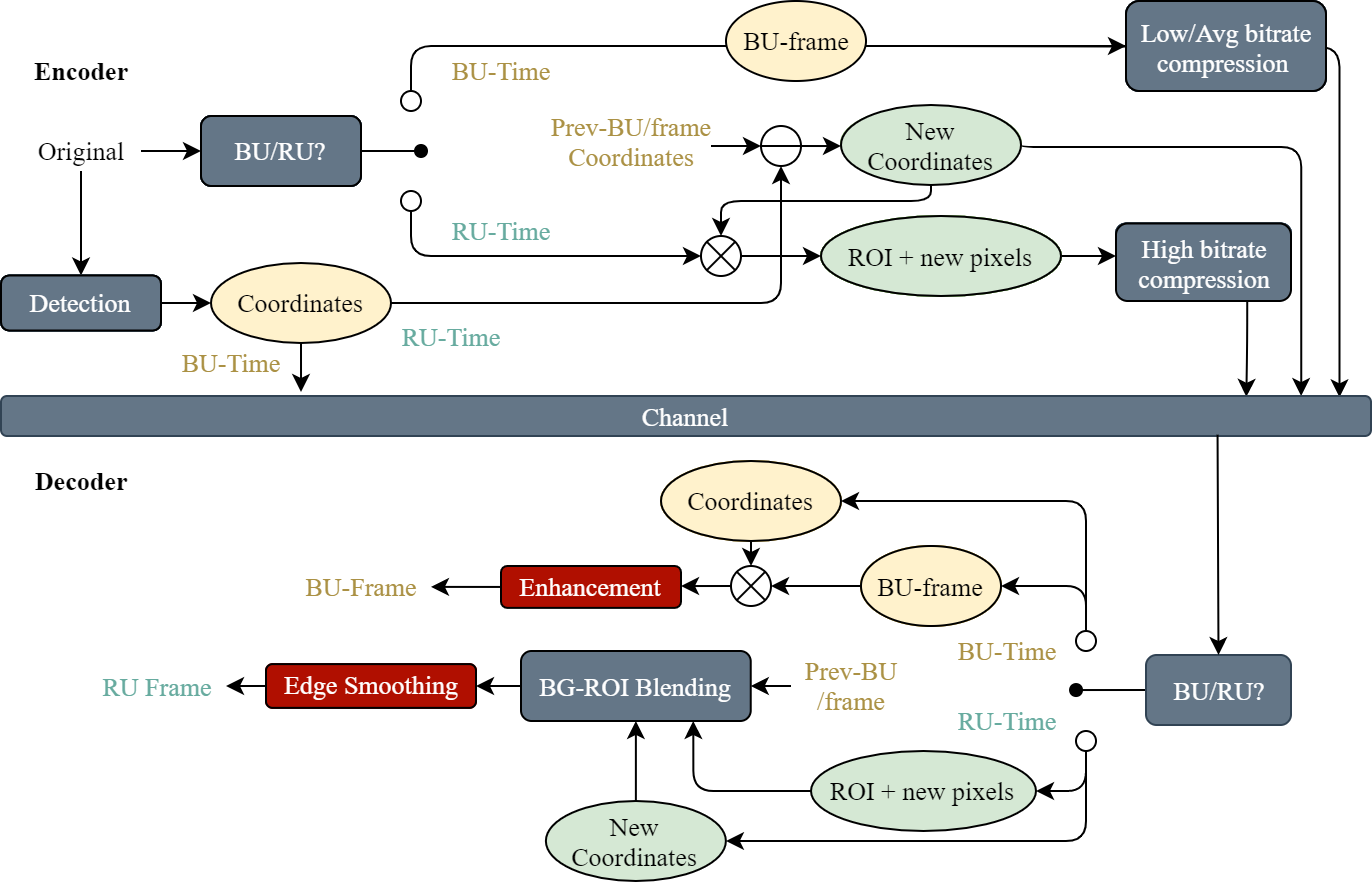}
\end{center}
   \caption{Overview of proposed RCLC framework.}
\label{fig:Overview}
\end{figure*}

With the spread of COVID-19, online meetings have become a must globally. Therefore, video streaming, which is the main instrument for a remote meeting system, is facing a huge demand for the video compression method improvement. Especially for high-resolution streaming video, existing methods are standing still because of the trade-off between the reconstruction quality and the specific real-time demand. To overcome that problem, effective video compression with a specific configuration is in need.

Current conventional video codecs usually perform a uniform compression for all-region in a video frame, however, in a remote meeting system, the background is usually pointless that most existing systems provide a blurring filter to hide it. Therefore, a Region-Of-Interests-based (ROI) codec, where the non-interest area will be compressed with a coarser quality than the ROI, is suitable for this problem. R. Delhaye \textit{et al.} \cite{8629050} introduced a QP-selection scheme for ROI on H.265 codec \cite{6316136}, this system extracts ROI based on a specific Kinect skeleton detection, which cannot be available for all normal users. L. Zhonglei \textit{et al.} \cite{8830443} announced a faster codec for specific airport cameras by sending the specific uniform information of background instead of the whole background, it can reduce a lot of bitrate, however, it results in a very bad visual quality. Then, L. Wu \textit{et al.} \cite{9208694} introduced a learning-based ROI compression framework that dynamically chooses the frame to update as background or do the background interpolation instead, however, this framework cannot perform in a streaming application manner.

To overcome the drawback of existing methods, in this work, we introduce an ROI-based join Conventional and Learning Compression framework (RCLC) that can support a real-time video streaming demand with 1080p resolution on a normal PC architecture. For the coming frame, we mark it as a background-updating (BU) or ROI-updating (RU) frame. BU frame is compressed with low quality while the ROI from RU-frame is compressed with high quality by the conventional codec to satisfy the real-time demand. Then, using learning-based methods, we can reconstruct the full RU frame on the decoder-end based on the BU-frame or previous frame. The experimental results show that our RCLC can reduce up to 32.55\% bitrate for the ROI region compared to H.265 video codec while supporting a real-time video streaming demand with 1080p resolution on a normal PC architecture. Our contribution is mainly three folds:

\begin{itemize}
\item We propose an ROI-based join Conventional and Learning Compression framework (RCLC).  It can be combined and extended with any existing/future learning-based methods and video codec to further improve its performance and adapt to any ROI demand.
\item Based on our framework, we experiment with several settings to further exploit its demand adaption ability.
\item Moreover, we propose two 1080p ROI-based testing sets with ROI is defined as a person. The first set is an online-meeting-related set, where the camera position is fixed indoor while the second one has some changes in camera angles over time and mostly outdoor scenes.
\end{itemize}

\section{Related work}
Existing conventional video codecs are well-known for their hand-crafted artifacts. For example, in the case of real-time video streaming over a narrow network bandwidth, the typical block-based video compression codecs such as H.264/AVC \cite{1218189} and H.265/HEVC \cite{6316136} get involve in those artifacts because of the large quantization parameter and coding unit size. Moreover, since those codecs usually perform the uniform compression, in which all regions have the same assigned parameters, the quality of Region-Of-Interests (ROI) is similar to the rest. Addressed this problem, \cite{8830443} have sent the meta-data instead of pixels values for the background while compressing the ROI with high quality. However, their final reconstructed frame is much like machine-generated with many visible blending edges since that meta-data is not enough to synthesize the ROI and the background differences. In our work, different from \cite{8830443}, we only send the meta-data that indicates the ROI region in the frame while repeatedly update the background information and using the learning-based methods to avoid the blending edges in the reconstructed frame.

Recently, several learning-based methods have been applied to improve ROI-based compression. \cite{8629050} used a specific Kinect device to extract the human skeleton information and do the ROI compression while assigning the quantization factors of H.265 according to the detected area. \cite{8943263}\cite{8855819} performed learning-based end-to-end ROI compression that learns and sends the ROI allocated map to the decoder. Meanwhile, \cite{9208694} introduced a foreground-background parallel compression for surveillance video, \cite{9208694} can remove a sufficient amount of bitrate for background compression by performing the background interpolation between two assigned background templates. However, this updating approach is not suitable to apply to video streaming tasks. Also, \cite{8943263}\cite{8855819} and \cite{9208694} require a lot of computation to perform the compression. In our work, all components and network has been carefully chosen to avoid the over complexity problem.

In collaborating the conventional and learning-based methods, we propose an ROI-based joint Conventional and Learning Compression framework (RCLC) that leverages the unused GPU with learning-based methods to improve the performance of ROI-based conventional compression. By defining ROI as a person, our RCLC uses the conventional codec to compress the non-important background as fast as possible while compressing ROI with high quality. Whereas, the learning-based methods are used as the ROI detector, background-ROI blender, and enhancing methods with a specific configuration. 

\begin{figure*}[t]
\begin{center}
\includegraphics[width=0.95\linewidth]{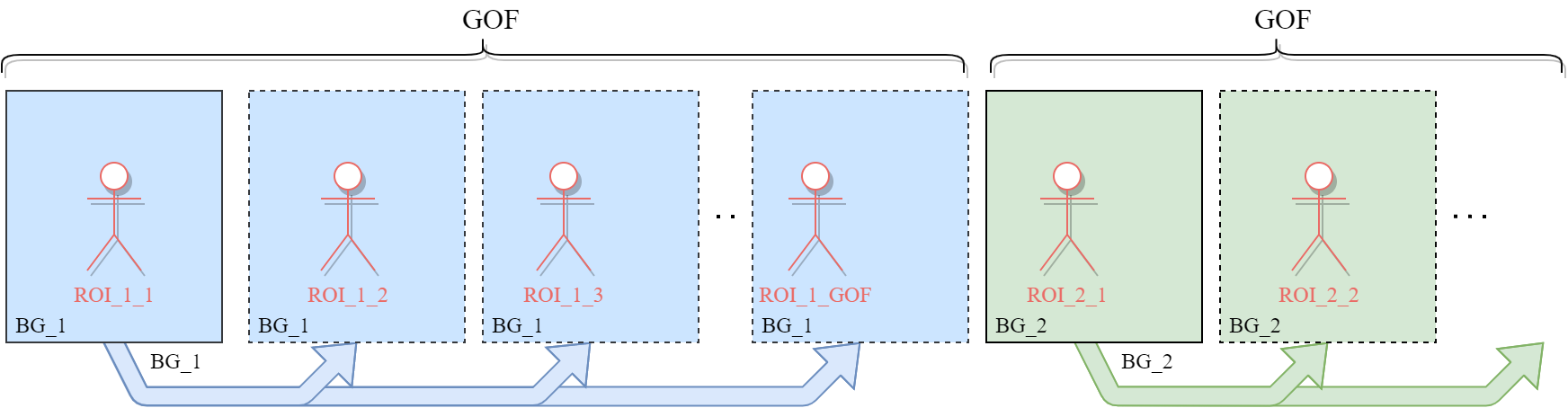}
\end{center}
   \caption{The concept of GOF and the concept of Background and ROI updating. BG, and ROI stand for background and Region-Of-Interests, respectively.}
\label{fig:GOF}
\end{figure*}

\begin{figure}[t]
\begin{center}
\includegraphics[width=\linewidth]{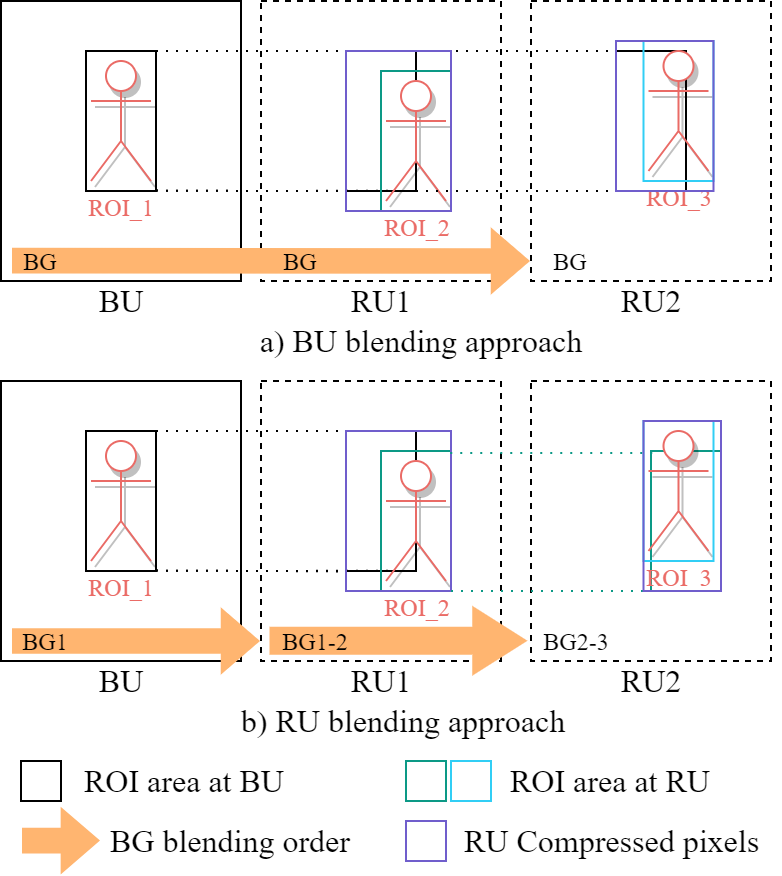}
\end{center}
   \caption{True compressed area for RU.}
\label{fig:ROICal}
\end{figure}

\section{The proposed RCLC framework}
\subsection{Overview of RCLC framework}
As shown in Figure \ref{fig:Overview}, for the incoming frame, we decide whether it is background-updating (BU) or ROI-updating (RU) frame based on our Group-Of-Frame (GOF) definition (see Section \ref{sec:GOF}). For BU, the whole frame is compressed by a codec that can satisfy the realtime-demand in collating to the hardware. Because BU’s purpose is to provide the background information, which is not much important, the codec is configured with a high quantization parameter to get very fast compression speed and low-bitrate. 

For RU, we have to recalculate the compressed area since we reused the previous background (see Section \ref{sec:GOF} and \ref{sec:ROICal}), only the new calculated ROI area is compressed with a low quantization parameter, although ROI area is usually much smaller than the whole frame, we still need to carefully choose the parameter in considering about compression speed. On the decoder side, we receive the information of the whole frame for BU and the ROI area for RU along with the position information. While the ROI-area of BU is enhanced, based on the received position information, ROI in RU is blended into previous reconstructed BU or previous frame and an edge smoothing network will filter-out the blended edges (see Section \ref{sec:enhance}).

\subsection{GOF and RU-RU}\label{sec:GOF}
Since background information is not so important in an online meeting, we need to limit the number of bits used for background transfers. Therefore, in our framework, Group-Of-Frame is defined as the period that background is updated, which means the frame at the start of GOF is chosen as BU (see Figure \ref{fig:GOF}). This definition is conducted based on our consideration of normal online meeting video, where the camera angle is rarely changed, so the background remains similar in a close range of visual similarity.

In each GOF, BU provides the background information for all the following RU. Whereas, the compressed ROI of RU is calculated not only based on the RU-frame but also based on the ROI position in BU or the previous frame (see Section \ref{sec:ROICal}). This manner is to avoid missing information when the ROI moving over frames.

\begin{figure*}[t]
\begin{center}
\includegraphics[width=0.9\linewidth]{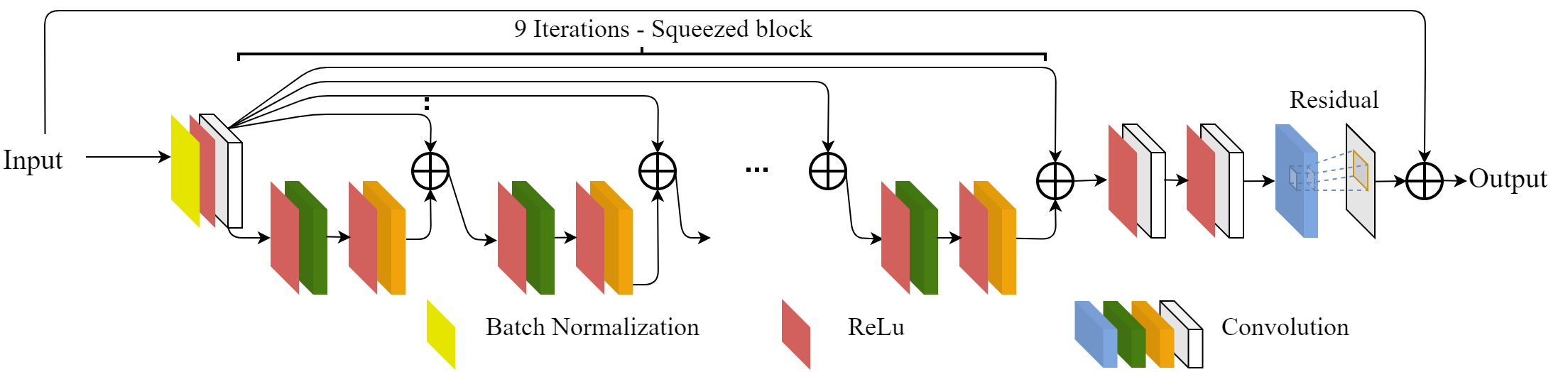}
\end{center}
   \caption{Edge smoothing network architecture.}
\label{fig:DRRN}
\end{figure*}

\begin{figure*}[t]
\begin{center}
\includegraphics[width=\linewidth]{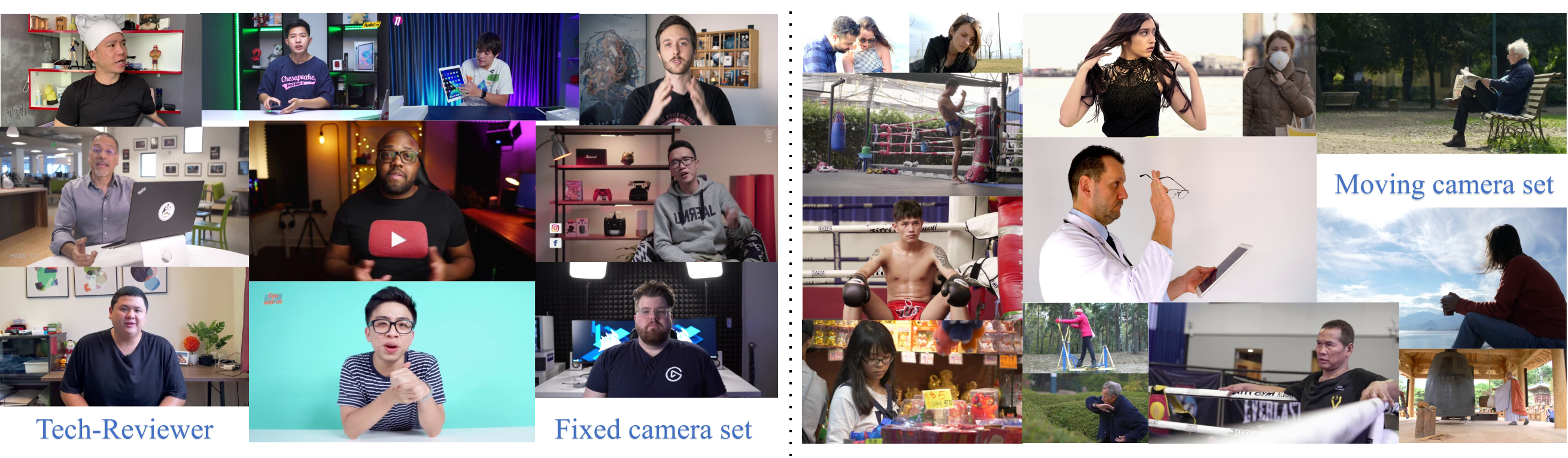}
\end{center}
   \caption{Some examples from proposed test sets.}
\label{fig:Dataset}
\end{figure*}

\begin{figure*}[t]
\begin{center}
\includegraphics[width=\linewidth]{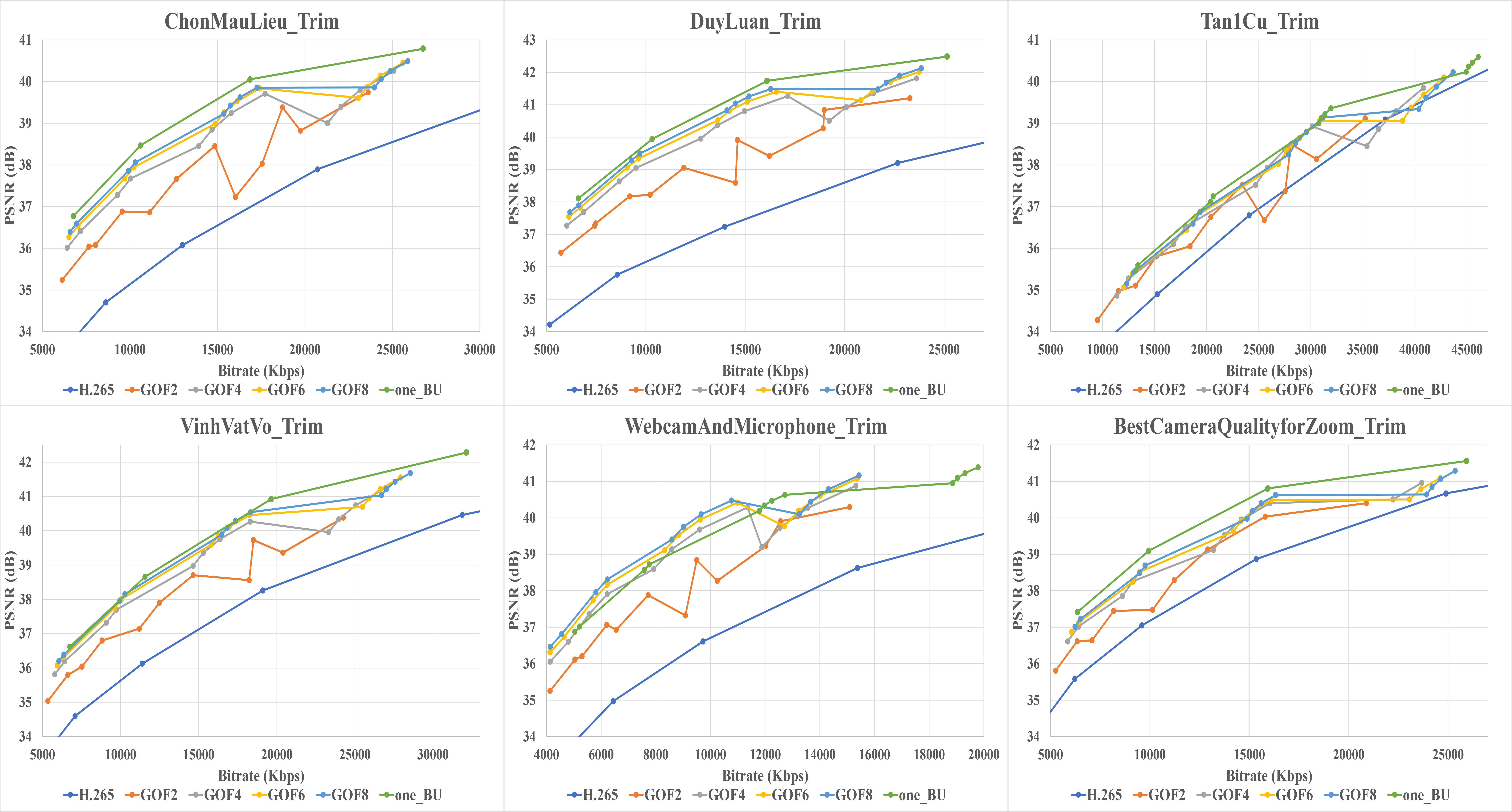}
\end{center}
   \caption{RD-curves on fixed-camera test set over difference GOF's size selection. Note that one\_BU means we use only the background of first frame for the whole video.}
\label{fig:RD-curve}
\end{figure*}

\begin{table*}[t]
\centering
\caption{The BD-rate reduction (\%) of our RCLC with GOF = 2 over H.265 codec on the ROI are of two proposed test sets.}
\label{tab:BDrate_GOF3}
\begin{tabular}{|l|c|l|c|}
\hline
\multicolumn{4}{|c|}{\textbf{Moving camera set}}                                                                                 \\ \hline
\multicolumn{1}{|c|}{\textbf{Sequence}} & \textbf{BD-rate (\%)} & \multicolumn{1}{c|}{\textbf{Sequence}} & \textbf{BD-rate (\%)} \\ \hline
Model\_36850                            & -15.33                & 180301\_07\_A\_HongKongIsland\_11      & -70.89                \\ \hline
190312\_24\_ParkVillaBorghese\_UHD\_002 & -55.78                & Woman\_23644                           & -42.04                \\ \hline
180301\_15\_A\_KowloonPark\_08          & -21.72                & 190111\_16\_MuayThaiTraining\_UHD\_01  & -16.78                \\ \hline
180626\_28\_BongeunsaTemple\_02         & -66.54                & Coffee\_20564                          & -38.71                \\ \hline
190111\_16\_MuayThaiTraining\_UHD\_06   & -34.34                & TrainingApparatus\_1087                & -34.31                \\ \hline
200323\_Coronavirus\_01\_4k\_014        & -25.77                & Doctor\_22704                          & -2.60                 \\ \hline
Couple\_19706                           & -8.96                 & 190111\_16\_MuayThaiTraining\_UHD\_12  & -59.81                \\ \hline
\multicolumn{1}{|c|}{\textbf{Average}}  & \multicolumn{3}{c|}{\textbf{-35.26}}                                                   \\ \hline \hline
\multicolumn{4}{|c|}{\textbf{Fixed camera set}}                                                                                  \\ \hline
\multicolumn{1}{|c|}{\textbf{Sequence}} & \textbf{BD-rate (\%)} & \multicolumn{1}{c|}{\textbf{Sequence}} & \textbf{BD-rate (\%)} \\ \hline
BestCameraQualityforZoom\_Trim          & -24.01                & LiveStreamingOrVideoConferencing\_Trim & -23.64                \\ \hline
ChonMauLieu\_Trim                       & -40.80                & NobodyKnew\_Trim                       & -24.07                \\ \hline
DuaLeo\_Trim                            & -34.40                & Tan1Cu\_Trim                           & -22.77                \\ \hline
DuyLuan\_Trim                           & -48.61                & VinhVatVo\_Trim                        & -35.21                \\ \hline
DuyTham\_Trim                           & -29.91                & WebcamAndMicrophone\_Trim              & -42.13                \\ \hline
\multicolumn{1}{|c|}{\textbf{Average}}  & \multicolumn{3}{c|}{\textbf{-32.55}}                                                   \\ \hline
\end{tabular}
\end{table*}

\section{Specific update interval}
\subsection{ROI calculation}\label{sec:ROICal}
To avoid specific hardware demand, we use the YOLO\_v4 method \cite{bochkovskiy2020yolov4} to perform the ROI detection task. YOLO\_v4 can do the detection by input a normal RGB camera frame. To meet the real-time with limited hardware resources, the ‘tiny’ version \cite{wang2021scaledyolov4}, which has a smaller weight-load, is used. Here, we define the ROI as one person.
For BU, the detected area is formed to two points of bounding boxes area, this information is used at the decoder to enhanced the ROI-area in BU. Furthermore, in the same GOF, RUs also use this information to recalculate their ROIs.

For RU, after performing normal detection as BU, since the object or ROI can move over time, a direct blending of ROI will lead to the missing pixels information. Therefore, by calculating the difference with the BU ROI bounding box in the same GOF, we can include the missing pixels into the new ROI  (BU\_blending in Figure \ref{fig:ROICal}.a). Furthermore, for large GOF size, where the number of missing pixels between the BU and RU increases due to the large movement of ROI, we introduce another updating rule. Instead of using BU as the background provider, we calculate the compressed area of current RU based on the previous frame, therefore, the distance of movement and the number of missing pixels are reduced, especially on very large GOF size (RU\_blending in Figure \ref{fig:ROICal}.b).  

\subsection{BU-RU enhancement}\label{sec:enhance}
Since there is a gap in the quantization parameter between BU and RU frames, the smoothness of visual quality over the sequences may be affected in the case of the limited bandwidth, at which we have to increase that gap. Hence, we have to perform an enhancement step for BU to catch up with the RU visual quality. However, since the background is not useful information, we only enhance the ROI area of BU on the decoder-end based on the received bounding box information.

While BU facing the quality gap with RU, the ROI inside of RU also facing that gap with the blended background. The blending approach usually leads to noticeable edges at boundary pixels. Therefore, we have to smooth those edges by inputting the decoded-ROI of RU plus the neighboring pixels of the blended background to an edge smoother.

We try to avoid using two separated model-weight sets for enhancing and edge smoothing because of the hardware on normal user PC consideration. Therefore, we use the deep recursive residual network \cite{8099781} as our enhancement model. This model is suitable for small GPU consumption with its recursive approach while easy to perform fine-tuning and avoid the overfitting problem. After training this network on the edge smoothing task, we fine-tune the network by squeezing the recursive residual block and input layer while training on the quality enhancement task (see Figure \ref{fig:DRRN}).

\begin{figure*}[t]
\begin{center}
\includegraphics[width=0.86\linewidth]{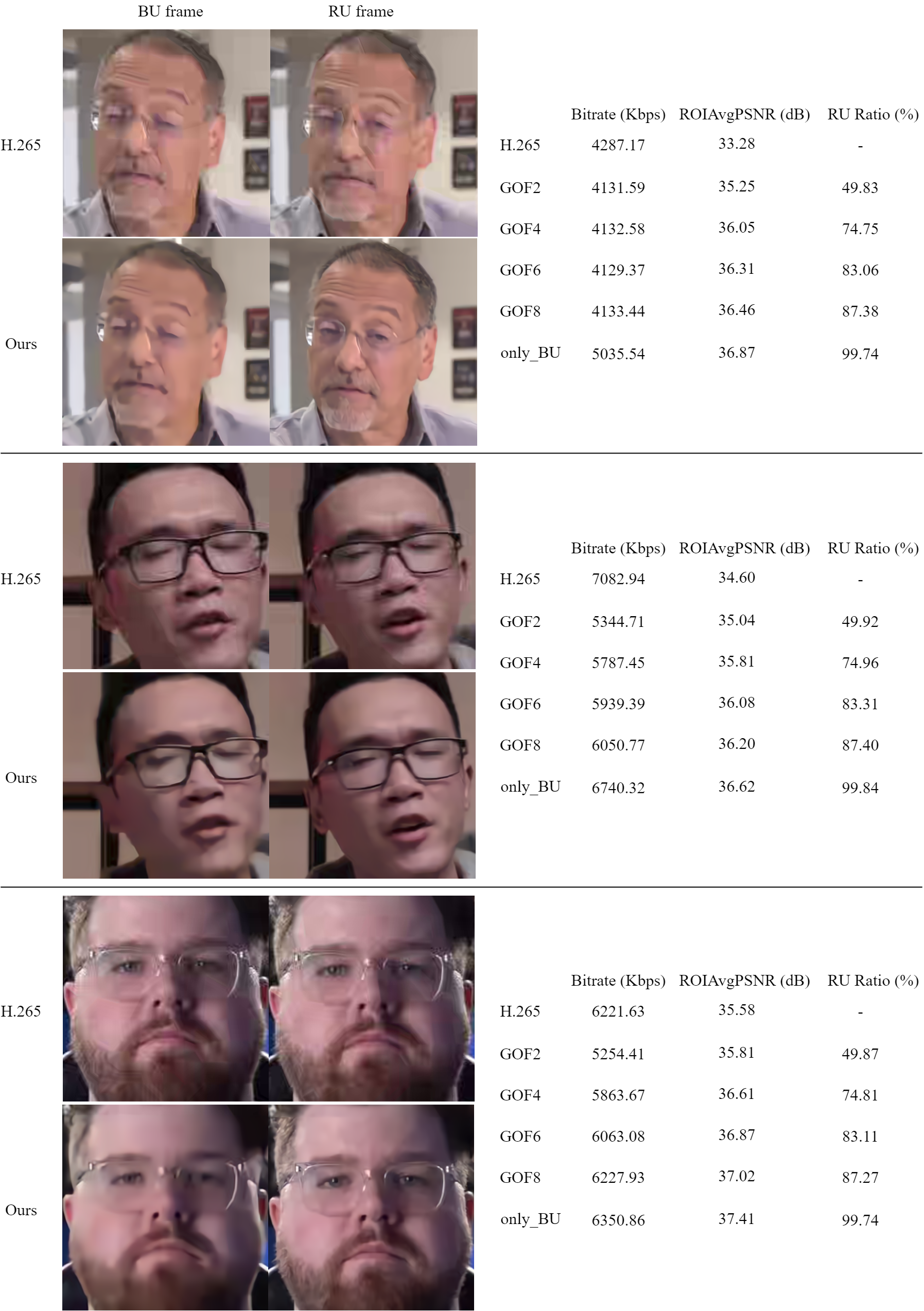}
\end{center}
   \caption{Some visual comparisons on fixed camera test set over H.265 codec.}
\label{fig:VisualResults}
\end{figure*}

\begin{table*}[t]
\centering
\caption{The BD-rate reduction (\%) of our RCLC over H.265 codec with two different blending approaches on fixed camera test set. Here, b\_ and r\_ denote for BU-blending and reconstructed RU-blending, respectively.}
\label{tab:b_r_compare}
\begin{tabular}{|l|c|c|c|c|c|c|}
\hline
\multicolumn{1}{|c|}{\textbf{Sequence}} & \textbf{bGOF4} & \textbf{rGOF4}  & \textbf{bGOF8} & \textbf{rGOF8}  & \textbf{bOne\_BU} & \textbf{rOne\_BU} \\ \hline
BestCameraQualityforZoom\_Trim          & -31.00         & \textbf{-31.43} & -34.31         & \textbf{-35.33} & \textbf{-39.87}   & -39.20            \\ \hline
ChonMauLieu\_Trim                       & -49.60         & \textbf{-49.74} & -53.42         & \textbf{-53.77} & -56.78            & \textbf{-57.27}   \\ \hline
DuaLeo\_Trim                            & -36.21         & \textbf{-36.71} & -42.94         & \textbf{-43.72} & -45.51            & \textbf{-46.25}   \\ \hline
DuyLuan\_Trim                           & -57.49         & \textbf{-57.70} & -60.98         & \textbf{-61.36} & -62.55            & \textbf{-64.40}   \\ \hline
DuyTham\_Trim                           & -37.33         & \textbf{-37.41} & -40.51         & \textbf{-40.95} & -40.94            & \textbf{-44.22}   \\ \hline
LiveStreamingOrVideoConferencing\_Trim  & -21.50         & \textbf{-21.87} & -23.97         & \textbf{-24.57} & -24.15            & \textbf{-27.43}   \\ \hline
NobodyKnew\_Trim                        & -29.58         & \textbf{-29.95} & -32.66         & \textbf{-33.31} & -36.47            & \textbf{-36.79}   \\ \hline
Tan1Cu\_Trim                            & -23.25         & \textbf{-23.56} & -23.69         & \textbf{-24.16} & -25.40            & \textbf{-25.07}   \\ \hline
VinhVatVo\_Trim                         & -43.40         & \textbf{-44.22} & -46.70         & \textbf{-48.16} & -46.93            & \textbf{-51.74}   \\ \hline
WebcamAndMicrophone\_Trim               & -52.42         & \textbf{-54.77} & -56.62         & \textbf{-59.59} & -51.34            & \textbf{-63.68}   \\ \hline
\multicolumn{1}{|c|}{\textbf{Average}}  & -38.18         & \textbf{-38.74} & -41.58         & \textbf{-42.49} & -42.99            & \textbf{-45.60}   \\ \hline
\end{tabular}
\end{table*}

\subsection{Time constraint calculation}
Based on our framework, the computation time of the system will be calculated as follows:
\begin{itemize}
\item For BU: Encoder time = max(Detection, Compression); Decoder time = Decompression + ROI Enhancement. 
\item For RU: Encoder time = max(Detection) + Re-ROI calculation + Compression; Decoder time = Decompression + ROI Blending + Edge smoothing.
\end{itemize}
By setting the H.265 codec with ultra-fast preset and all\_intra configuration, our system can get framerate higher or equal to 60fps for all components on 1080p video.

\section{Results and Comparison}
\subsection{Experimental Setting}
Our experiments were conducted on an NVIDIA RTX 2080Ti GPU while an Intel Core i7-8700K CPU was used to perform non-GPU tasks. For training our enhancement/edge smoothing network, we collected 20 videos with CIF and 720p resolution from \cite{Xiph}, all frames were extracted and used. We used compressed frames as input for the enhancement task. For the smoothing task, the ground-truth ROI was blended to the compressed background to form a combination input. All frames were ROI-centered cropped with the size of 512x512 for a fixed training size. We implemented our proposal using the PyTorch\cite{pytorch} framework. We used Adam\cite{kingma2014adam} optimizer and started the edge smoothing task learning with a learning rate of 1e-04, then terminated the training after 20 epochs. Next, we perform the enhancement task training with layers freezing (see Section \ref{sec:enhance}) and learning rate of 1e-05 then terminated after 10 epochs.         

For testing, since there is no available online meeting related non-compressed dataset with 1080p resolution, we conducted two new test sets. The first set is for the online meeting scheme, where the camera angle is fixed (fixed camera set/tech-reviewer set) and the second one has the viewpoint that changes over time (moving camera set). We collected 24 videos with 4K resolution from YouTube\cite{ytb} and Videvo\cite{vivo}, then down-scaled them to 1080p using a bicubic operator to remove the compression artifacts. Especially, for the fixed camera set, because there is almost no available 4K online meeting video, we collected the sequences from tech-reviewers, which also satisfies the scheme requirement. Figure \ref{fig:Dataset} shows a brief view of our proposed test sets. In our experiments, we use PSNR and Bjontegaard-delta (BD)-rate \cite{bjontegaard2001calculation} metrics to evaluate our results. 

\subsection{Comparison}
We compare our RCLC with H.265 codec\cite{6316136} over two proposed test sets. Note that, we also use H.265 as our anchor in this comparison. For H.265, QPs equal to 32, 37, 42, 47 which ensure the 60fps compression. For our RCLC, QPs for ROI are 22, 27, 32, 37, while QPs for background are 32, 37 (for only ROI QPs = 22, 27), 42, 47. Except for the \textbf{Blending approach selection} comparison, we use the BU\_blending approach (see Section \ref{sec:ROICal}) for all comparisons because it can cover the smallest GOF size, which is suitable for moving camera cases.

\textbf{Bitrate reduction compare to H.265}. As shown in Table \ref{tab:BDrate_GOF3}, we first compare our RCLC using our basic GOF = 2 with H.265 codec on the ROI are of two proposed test sets. We can see that our RCLC got better results for ROI quality than H.265. In particular, our RCLC can gain up to 35.26\% BD-rate reduction over H.265 on the moving camera set and 32.55\% on the other. In the best case, our RCLC can reduce up to 70.89\% RD-rate and 2.6\% for the worst case of Doctor\_22704 sequence, where the ROI area is bigger or equal to two-third of frame size. It is worth mentioning that our RCLC can achieve this result while keeping the compressing speed intact from H.265 by leveraging the unused GPU hardware. 

\textbf{GOF selection}. Figure \ref{fig:RD-curve} shows several RD curves of our RCLC over different GOF selections on the fixed camera set. We can see that when the camera angle is fixed, by increasing the GOF, RCLC can get better performance. And in an ideal situation, where the camera is fixed and the person ratio does not change much during the meeting, we can set the GOF equal to full sequence (one\_BU), which means we only need to send the background only one time. 

\textbf{Visual results}. We compare several compressed frames with H.265 in Figure \ref{fig:VisualResults} . When looking into the cropped ROI part, we can see a lot of noise and block artifacts from H.265 compression. While with smaller transferring bits, our RCLC can get higher-quality visual in texture and smoothness for the ROI with all GOF. In our RCLC framework, BU may have worst quality than RU, however, by increasing the GOF's size, the RU ratio also increases which leads to the higher average PSNR value for the ROI area along with the sequence's frames. Hence, in the ideal case, where only the first frame is BU, all remained frames will have much better visual quality compared to H.265.

\textbf{Blending approach selection}. We further exam two blending approaches in Section \ref{sec:ROICal}. The results of BD-rate reduction over GOFs = 4, 8, and one\_BU on fixed camera set are tabulated in Table \ref{tab:b_r_compare}. Here, b\_ and r\_ denote BU-blending and RU-blending approaches, respectively. We can see that the RU-blending approach gets better performance than the BU-blending approach for most cases. By increasing the GOF size, the performance distance increases from 0.56\% at GOF4 to 0.91\% at GOF8 and 2.61\% at the ideal one\_BU case. The performance gain comes mainly from the smaller compressed area because of the minimum motion between two consecutive frames. This result demonstrates that the RU-blending approach is much suitable for stable sequences which have fixed camera angle over time.          


\section{Conclusion}
This work presents a light ROI-based joint conventional and learning compression framework for video streaming. With our idea of interval background updating and reusing, we can reduce a sufficient amount of background transferring bit. Furthermore, our specific training procedure also reduces the required GPU storage twice for storing the learning-based enhancement network. Experiment results show that our RCLC outperforms the H.265 codec and gains up to 32.55\% BD-rate reduction while having a competitive compressing speed. Moreover, our framework is flexible that can be applied to any ROI problem in the future.

{\small
\bibliographystyle{ieee}
\bibliography{egbib}
}

\end{document}